# Significance of Medium-Wave AM Radio Broadcasting for Enhanced Disaster Resilience in Japan: A Case Study in the Kanto Region and Fukui using Nonpowered HOOPRA

Eiichi Shoji

*Department of Mechanical and System Engineering, Advanced Materials Innovation & Monozukuri Lab, University of Fukui, 3-9-1 Bunkyo, Fukui 910-8507, JAPAN*
*Email: shoji@u-fukui.ac.jp*

**Abstract**
On the occasion of centenary anniversary of the Great Kanto Earthquake and commencement of radio broadcasting in Japan, this study reiterates the paramount importance of medium-wave (MW) AM broadcasting in safeguarding public safety and security. Utilizing the electromagnetic principles of MW, the author has earlier developed hoop-shaped radio (HOOPRA), which is a battery-free sustainable radio receiver. This study aims to determine the maximum achievable reception distance with HOOPRA for broadcasts from public stations, such as Nihon Hoso Kyokai (NHK) JOFG (927 kHz, 5 kW) in Fukui, and NHK JOAK (594 kHz, 300 kW), and JOAB (693 kHz, 500 kW), in the Kanto region. The significance of the findings in this study is that approximately 38 million individuals in the Kanto region, residing within an 80 km radius of JOAK or JOAB, can access broadcasts using only the energy of radio waves with HOOPRA. Additionally, ~0.4 million people in Fukui, within a 15 km radius of JOFG, can potentially be recipients of the broadcast. Given that most transmitting stations operate at 5 kW nationwide, HOOPRA can be effectively utilized within a 15 km radius of each station. Moreover, these outcomes validate the efficacy of HOOPRA as a radio receiver and provide valuable insights into the global potential applicability of MW AM radios. Furthermore, the current investigation underscores the need to reevaluate the significance of terrestrial MW broadcasting as a vital source of emergency information, especially in the event of anticipated natural disasters, such as the predicted Nankai Trough Earthquake.

**Keywords**: Disaster-prevention radio, Medium-wave AM broadcasting, Nonpowered radio, Sustainable radio, HOOPRA

## 1. Introduction

The seismic catastrophe known as the Great Kanto Earthquake occurred on September 1, 1923, and its epicenter was located on the coast of the Miura Peninsula in Kanagawa Prefecture. This earthquake with an estimated magnitude of M7.9 [1] inflicted widespread devastation in regions such as Tokyo, Kanagawa, Chiba, and Saitama prefectures. The toll exacted by this disaster stands as a tragic testament to its severity, with over 0.1 million lives lost or reported missing, cementing its position as one of the most catastrophic events in Japanese history [1-3]. The collapse and crushing of houses directly caused the demise of ~10,000 individuals, while the ensuing conflagration claimed the lives of ~90,000 people. Amid the difficulty of rescuing victims and disseminating information, the propagation of rumors throughout the affected areas exacerbated the already tragic casualties and incidents [2]. Multitudes of lessons have been learned from the disaster, including the imperative of establishing a swift and dependable information network during the disaster [3].

This catastrophic event exposed the societal and governmental challenges that were prevalent during that period in Japan [1-3]. In response, it prompted a series of transformative measures aimed at strengthening disaster preparedness and reevaluating urban planning. Among these groundbreaking changes was the commencement of radio broadcasting in Japan [3], which emerged as a vital conduit for swift and reliable information dissemination. Prior to the Great Kanto Earthquake, radio broadcasting has yet to be established in Japan. Notably, the Ministry of Communications had been actively promoting the establishment of a broadcasting system in the waning years of the Taisho period, immediately before the onset of the turbulent era preceding the calamitous event of the Great Kanto Earthquake [4,5]. Deliberations on the introduction of radio broadcasting were underway, even on the eve of the earthquake, poised to herald the advent of this technology in Japan [4].

The radio, an innovation utilizing wave-based signal transmission, was first invented by Marconi in 1895 [5-11]. Its initial applications was realized using Morse code wireless communication, and it successfully achieved the wireless transmission of audio signals by 1900 [5,7,9,10,12]. This breakthrough propelled the global proliferation of radio broadcasting technology, culminating in the establishment of the world's first commercial radio station, KDKA, in the United States in 1920 [5,7,9,10,13]. In Japan, the inaugural radio broadcast started on March 22, 1925, conducted by the Tokyo Broadcasting Station (JOAK), which now operates as Nihon Hoso Kyokai (NHK), had only a temporary broadcasting station in Shibaura then. Under the purview of the Ministry of Communications, Tokyo Broadcasting Station (JOAK), Nagoya Broadcasting Station (JOCK), and Osaka Broadcasting Station (JOBK) were granted official approval [5-10,14,15]. Crystal radios predominated as the primary receivers during this era, as foreign vacuum tube radios were prohibitively expensive and had yet to permeate the general populace. In the 1930s, radio broadcasts primarily pioneered by NHK started targeting households and gained momentum before firmly establishing radios as a ubiquitous fixture of home entertainment [5-10,15].

The NHK Tokyo Radio First Broadcast (JOAK), which emanates from Shobukuki in Saitama, operates at a transmission output of 300 kW, while the NHK Tokyo Radio Second Broadcast (JOAB) stands as one of the nation's largest AM radio broadcasts, boasting a transmission output of 500 kW [16]. The Kanto region, encompassing Tokyo and six prefectures (Ibaraki, Tochigi, Gunma, Saitama, Chiba, and Kanagawa), is home to approximately 35% of Japan's total population, amounting to 43.53 million people [17]. In times of disaster, the NHK radio broadcasts from Shobukuki play a vital role for residents in the Kanto and Koshinetsu regions.

For over a century, the method of receiving radio signals solely through the electromagnetic energy of medium-wave (MW) broadcasting has been known, exemplified by germanium or crystal radios. Although these radios can function without an external power supply, their requirement for a sizable external antenna hindered their widespread adoption as portable radios. To the best of the author's knowledge, no scholarly reports have demonstrated the development of an exceptional portable nonpowered radio receiver or provided insights into its reception range. Addressing this gap, the author has been engaged in the advancement of an energy harvester that obviates the need for an external antenna, accomplished through the evolution of a multiloop antenna and circuitry enhancements to capture MW electromagnetic waves with heightened sensitivity [18,19]. Thus, the author created a battery-free radio receiver known as hoop-shaped radio (HOOPRA), which neither requires an external antenna nor relies on conventional batteries [18-20]. Serving as a sustainable radio solution, HOOPRA exhibits exceptional portability, lightweight construction, flexibility, and user-friendly functionality.

The primary objective of this study is to elucidate the maximum distance at which the contents of radio can be broadcasted. For e.g., NHK Fukui Radio First Broadcast (JOFG) in Fukui and JOAK and JOAB in the Kanto region remain discernible when employing HOOPRA. These findings will subsequently facilitate the performance evaluation of HOOPRA as a disaster preparedness radio. JOFG operates at a transmission power of 5 kW, a typical output for local prefectural broadcasting station. Thus, these obtained findings using JOFG will be pivotal in exploring nationwide applicability. Furthermore, the reception performance findings obtained from JOAK and JOAB enable an estimate of the approximate population within the Kanto region that is capable of receiving these broadcasts. Additionally, considering that JOAB in Sapporo (JOIB), Akita (JOUB), and Kumamoto (JOGB) share a transmission power of 500 kW, and JOAK in Osaka (JOBB) has the same 300 kW power output as of JOAK, the reception insights garnered from JOAB and JOAK serve as valuable considerations when assessing the efficacy of HOOPRA for individuals residing in these regions.

With these outlined research objectives, the author seized the momentous occasion—the 100th memorial anniversary of the Great Kanto Earthquake and anniversary of the commencement of broadcasting—to reaffirm the invaluable lessons learned from significant disasters and underscore the paramount importance of terrestrial radio broadcasting in addressing this pertinent issue. Herein, the author will discuss upon the significance of radio broadcasting as the most dependable and trustworthy information source, drawing upon lessons derived from past calamities and elucidating the critical infrastructure that will be crucial during future disasters in Japan. From a practical standpoint, the author will highlight the performance of HOOPRA as a sustainable radio receiver for disaster prevention. Notably, the Nankai Trough Earthquake holds a 70%–80% probability of occurring within the next 30 years [21]. The contents of this paper are poised to provide invaluable insights for various impending natural disasters while concurrently serving as a catalyst to enhance awareness regarding disaster-prevention efforts.

## 2. Method

### 2.1. Development of HOOPRA

HOOPRA, a nonpowered radio receiver, harnesses the energy of electromagnetic waves from MW AM broadcasts as a radio-wave energy harvester. This radio significantly differs from conventional germa or crystal radios crafted with artistic coil winding. However, it offers practical sensitivity, remarkable portability, a lightweight design, and robust construction achieved through the implementation of specially devised wire materials that eliminates the need for an external antenna. The absence of battery requirements eliminates laborious maintenance tasks. It can be conveniently stored in stockpile warehouses or emergency carry-out bags, readily available as a disaster-prevention radio at any instant. Concerns regarding battery depletion, deterioration, or leakage during storage, often encountered with battery-operated radios, are entirely mitigated. Furthermore, it emits no winding noise, a significant advantage in nerve-wracking evacuation shelters teeming with evacuees, and spares users the inconvenience of hand-cranking the radio. Moreover, worries related to sunlight availability, light source availability, charging time, or degradation of built-in secondary batteries found in solar-powered radios is dispelled by embedding a multiloop antenna. HOOPRA incorporates a flexible hoop-shaped design with a diameter of, for example, 60 cm (HOOPRA-60) (Figure 1A), housing a multiloop antenna. Instead of connecting HOOPRA to a single-ear earphone, users have the option to connect it to a smartphone and listen to terrestrial AM radio broadcasts with standard earphones. Notably, HOOPRA does not deplete the smartphone's battery. Given the ubiquity of smartphones today, combining HOOPRA with a smartphone enables the reception of AM broadcasts through terrestrial waves directly on the smartphone (Figure 1B). With the installation of a specific application on the smartphone, users can even record broadcasts during a disaster. In the rolled-up state, the diameter of HOOPRA reduces to one-third of its original size, rendering it a durable radio that can be conveniently mounted on a wall near the entrance of a house. Moreover, it can be utilized during disasters or carried in a bag for facile transport.

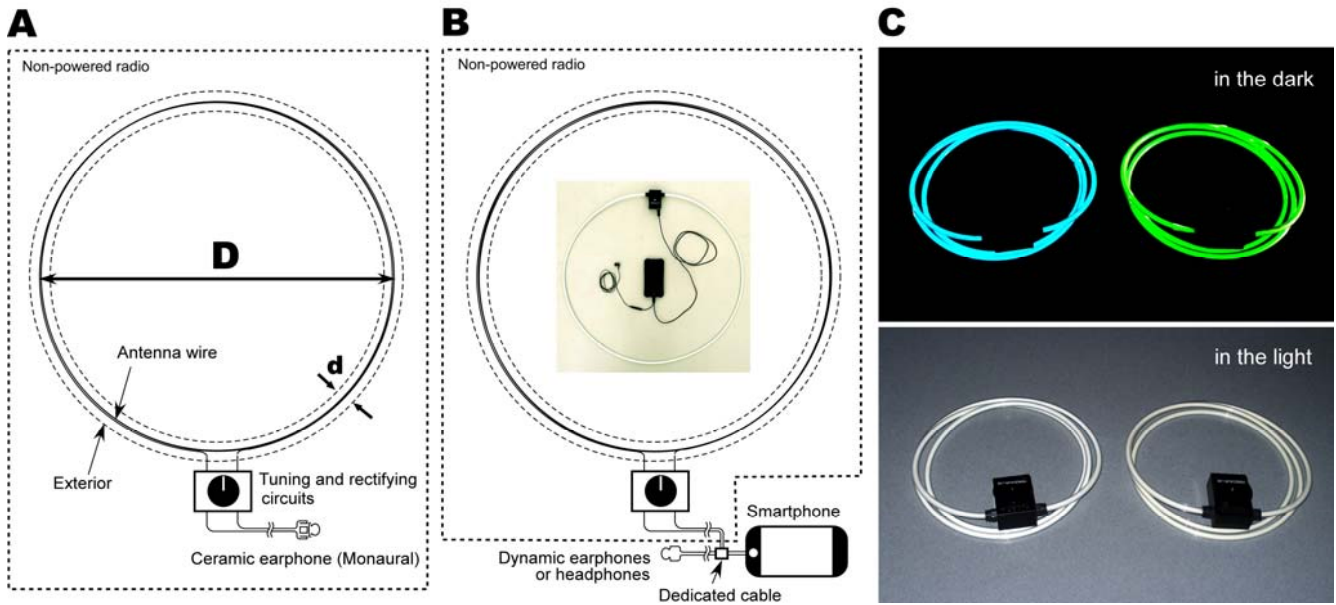

Figure 1   A: Schematic of nonpowered radio HOOPRA. HOOPRA-60: D = 60 cm, HOOPRA-80: D = 80 cm, d = 8 mm. B: Diagram of smartphone connected to HOOPRAs with a dedicated cable and the image of connected state (center photo). C: HOOPRA with highly bright phosphorescent material integrated into housing: top: glowing in the dark and bottom: an illuminated scene.

## 2.2. About the Field Experiments

The author's laboratory being located in Fukui, the broadcasting from JOFG (927 kHz, 5 kW, Geba Town, Fukui City, Fukui Prefecture) station was examined to verify the maximum distance over which the broadcast content remains discernible. Preliminary findings narrowed down the endpoints to 11 km at Nishiyama Park (Sabae City, Fukui Prefecture), 13 km at Sundome Fukui (Echizen Town, Fukui Prefecture), and 15 km at Murakuni Park (Echizen Town, Fukui Prefecture) to establish the practical limit of reception distance.

To assess the reception capability in the Kanto region, broadcasting from JOAK (594 kHz, 300 kW, Shobukuki, Kuki City, Saitama Prefecture) and JOAB (693 kHz, 500 kW, Shobukuki, Kuki City, Saitama Prefecture) stations were examined to determine the maximum distance at which the broadcast content remains recognizable. Preliminary findings narrowed down the locations to 60 km at Kita Hassaku Park (Yokohama City, Kanagawa Prefecture), 70 km at Kodomo Nature Park (Yokohama City, Kanagawa Prefecture), 70 km at Hikichidai Park (Yamato City, Kanagawa Prefecture), and 80 km at Kanazawa Zoo (Yokohama City, Kanagawa Prefecture) to determine the practical reception distance limit.

**2.3. Performance of HOOPRA in Disaster Drills**

During the comprehensive disaster-prevention drill held in Fukui City at the end of June 6, 2022, the author assessed the performance of the HOOPRA-60 prototype as a disaster-prevention radio. The performance of HOOPRA was evaluated using JOFG. The designated place of refuge was Nisshin Elementary School in Fukui City, which is approximately 5 km from JOFG in a straight line.

**2.4. About the Equipment Used**

During the field trials, HOOPRA-60 and HOOPRA-80 were utilized by the author. The compact size of HOOPRA-60 makes it suitable for easy mobility, allowing both hands to be free in a disaster scenario. To enhance nighttime visibility, the author has incorporated a composite of luminous material having ultrahigh brightness onto the flexible body of HOOPRA-60 [20]. The material selected for the ultrahigh brightness phosphorescent component was LumiNova ($SrAl_{12}O_4$:Eu, Dy, $Sr_4Al_{14}O_{25}$:Eu, Dy, Nemoto & Co., Ltd, Tokyo, Japan). To assess the field reception capability, the author measured the induced voltage at the resonance frequency of the LC resonant circuit using an alternating current (AC) voltmeter (VT-171, KENWOOD Corp., Tokyo, Japan), as well as the rectified induced voltage with a data logger (GL-980, GRAPHTEC Corp., Kanagawa, Japan). HOOPRA-80 was used for connecting the instruments. Furthermore, the author measured the alternating consumption current of a monaural ceramic earphone using an AC voltmeter through a resistance circuit. For HOOPRA, a germanium diode of IN34A (DO-7) was utilized, with the possibility of using an equivalent diode. Note that the reason for the preference for germanium diodes with low $V_f$ will be discussed in a future publication. For reception, the author employed a monaural ceramic earphone (7 kΩ at 1 kHz, Model: ZHW-CRE-7K-G, Yutoku Denshi Ltd., Nagasaki, Japan). As for the smartphone connected to HOOPRA, an iPhone SE3 (Apple Inc., California, USA) was utilized, along with a dedicated cable connecting the smartphone's lightning and HOOPRA audio port.

## 3. Results and Discussion

**3.1. Radio Broadcasts Resilient to Power Outages and Lessons Learned from Past Disasters**

According to a report evaluating various media usage during the Great East Japan Earthquake on March 11, 2011, AM radio received the highest assessment at 60.1%, followed by FM radio, during the earthquake [22]. Radio exhibited notable immediacy compared to television and social networking services (SNS). However, the utilization data for media (e.g., radio, television, and SNS) in these statistics does not account for situations during power outages. In reality, power outages frequently occur simultaneously with disasters such as typhoons and earthquakes, thereby rendering televisions and internet unusable, as observed during recent disasters. For instance, the Hokkaido Iburi Eastern Earthquake occurred at 3:07 a.m. on September 6, 2018, causing a significant blackout immediately after the earthquake. The entire Hokkaido region, encompassing approximately 2.9 million households, experienced power failure that required two days for restoration. Similarly, in Chiba Prefecture, which was directly hit by Typhoon 15 on September 9, 2019, an unusually prolonged power outage persisted for approximately 16 days, affecting 0.93 million households.

Moreover, on March 16, 2022, an earthquake struck Miyagi and Fukushima Prefectures at 11:34 p.m., leading to the shutting down of a power generation facility. The subsequent blackout affected ~2.1 million households in nine prefectures in the Kanto region at 00:05 h on March 17, 2022. This was the second-highest blackout experienced in Japan since the Great East Japan Earthquake. Historical examples have demonstrated that during such power outages, the usage of television, internet, SNS, and mobile phones can be compromised depending on the circumstances. In natural disasters, it is crucial to take action for safeguarding lives from tsunamis and floods. Although the disaster administration radio system, including simultaneous broadcasting via outdoor loudspeaker stations, provides information on life-threatening emergencies, the audibility of loudspeaker announcements can be challenging during heavy rain, thereby hindering evacuation. Thus, a reliable approach of accessing emergency information must be identified to consider the most effective method of obtaining timely and trustworthy information during a disaster. When power is disrupted and internet becomes unreliable, terrestrial radio broadcasting emerges as one of the few available sources of information. Although internet radio has gained popularity among certain individuals owing to the widespread usage of smartphones, it is unreliable during disasters and power outages as internet connectivity cannot be ensured. Power outages often occur in conjunction with disasters, rendering internet radio, SNS, and even phones unusable if the internet and routers are unavailable due to the power outage.

Furthermore, internet radio such as NHK poses limitations in providing specific local information during a disaster. Understanding the relationship between the target broadcasting area and hosting broadcasting station is crucial. For instance, NHK is broadcasted to each target area through eight broadcasting stations located across the nation (as of 2023.5). As the prefectural broadcasts available through terrestrial radio receivers differ from the internet radio, a user receives broadcasts on common information within the target broadcast area. For example, NHK's internet radio in Fukui broadcasts information through Chukyo Wide Area Broadcasting, which is managed by NHK Nagoya First

Broadcasting Station (JOCK). However, to obtain detailed information about the surroundings of Fukui City, a radio receiver capable of picking up JOFG's terrestrial waves is necessary. Even if infrastructure development enables the simultaneous provision of nationwide prefectural broadcasts and internet broadcasts to the public, the uncertainty of internet during disasters means that internet radio cannot be relied upon as a primary source of information. Instead, internet radio should serve as a supplementary measure during disaster.

### 3.2. Characteristics of MW AM broadcasting

FM radio broadcasts employ ultrashort waves (USW), whereas AM radio broadcasts use MW in radio broadcasting. The propagation properties of these waves differ significantly. Specifically, MW offers a distinct advantage over USW in propagating beyond line-of-sight distances due to the characteristics of ground and ionospheric waves. From a technical perspective, MW radio broadcasts, with their lower frequency, can generate higher power output from the transmitter tube. This technical aspect, combined with the physical phenomena involved, enables MW AM radio broadcasts to reach wider areas, including mountainous regions, beyond line-of-sight distances compared to ultrashort-wave FM radio broadcasts. FM radio broadcasts operate within line-of-sight distances through direct waves. Although FM radio broadcasting is becoming more community-oriented, its service area is restricted. In the event of large-scale disasters, multiple FM radio stations within the same area may be affected simultaneously. However, MW broadcasts can reach these areas from far outside the disaster zone, rendering AM radio as a more reliable medium in disaster scenarios.

AM radio broadcasting and reception primarily rely on the magnetic field component of radio waves, while FM radio utilizes the electric field component. The magnetic field component of AM radio waves tends to be attenuated by steel materials found in buildings with reinforced or steel structures, which can make reception challenging depending on the internal structure of the building. However, buildings with minimal magnetic field shielding can still receive AM broadcasts effectively over a wide range. Owing to the long wavelength of MW, it can even penetrate non-ferrous metal walls, despite the presence of buildings or mountainous areas in the path of the radio wave. These exceptional characteristics of MW broadcasting, coupled with Japan's radio regulations, can facilitate the ultrahigh-power transmission of 500 kW; hence, AM radio broadcasting is established as an essential infrastructure. The wide-area propagation characteristics of MW broadcast waves align with the principles of radio-wave propagation. Considering the aforementioned factors, MW AM radio broadcasting holds immense value as a reliable technology for propagating information during frequent natural disasters in Japan.

### 3.3. Field Experiments in Fukui

The author constructed HOOPRAs based on previous reports [19]. For HOOPRA-60, a transparent polyurethane tube was used as the flexible outer shell, and a super-high-luminance phosphorescent material was mixed with silicone resin and encapsulated in the tube. Detailed information regarding radio-wave reception within a 10 km radius will be addressed in a separate paper on MW wireless power generation. Initial findings indicated that the broadcast waves from JOFG could be received within a 10 km radius using HOOPRA-60, prompting an investigation into the limit distance for comprehending radio broadcast content. The author examined Nishiyama Park at the 11 km point, Sundome Fukui at the 13 km point, and Village National Park at the 15 km point. At Nishiyama Park (11 km), the induced voltage of HOOPRA-80 against JOFG was 60 mV, and the author could clearly understand the broadcast with HOOPRA-60. Similarly, at Sundome Fukui (13 km), the induced voltage of HOOPRA-80 against JOFG was 40 mV, and the broadcast content was intelligible with HOOPRA-60.

Furthermore, at Village National Park (15 km), the induced voltage of HOOPRA-80 against JOFG was 20 mV, and while the broadcast content was barely recognizable with HOOPRA-60, it was clear with HOOPRA-80. At this point, the AC voltage of the LC resonance circuit without the monaural ceramic earphone connected to HOOPRA-80 was measured to be 140 mV, and the alternative current upon connecting the earphone was approximately 5 μA. These findings will be valuable for enhancing reception capabilities in future research. The field trials led to the conclusion that the reception limit for HOOPRA-60 and HOOPRA-80 is approximately 13 km and 15 km, respectively. If the induced voltage with HOOPRA-80 is around 20 mV or higher, HOOPRA-60 and HOOPRA-80 can be employed for listening to JOFG broadcasts.

Analyzing the population within 11 km, 13 km, and 15 km radii of JOFG using municipal data from the 2020 Japanese census conducted by the Statistics Bureau of the Ministry of Internal Affairs and Communications [23], the approximate figures are 0.29 million (0.12 million households), 0.35 million (0.14 million households), and 0.4 million (0.16 million households), respectively. These findings are significant as they indicate that around 0.4 million people in Fukui, residing within 15 km of JOFG, can listen to JOFG broadcasts with HOOPRA-80. As such, JOFG is an example of numerous 5 kW class radio broadcasting stations nationwide. Therefore, in many cities and towns hosting stations with a transmission power of 5 kW, HOOPRA-60 and HOOPRA-80 can be utilized within a radius of approximately 13 km and 15 km from each station, respectively. MW broadcast waves primarily propagated as ground waves, such that the coverage distance may vary based on the geographical shape and geological properties.

During the comprehensive disaster-prevention training conducted in Fukui City, the author assessed the practicality of HOOPRA-60 by carrying it along. At the training site, the measured alternative current upon connecting the earphone was approximately 150 µA, and the broadcast was remarkably clear and understandable. Despite the presence of loud noises from specialized vehicles and pump trucks, evacuation announcements through loudspeakers, the murmuring of people, and the noise of helicopters, the JOFG broadcast could be comprehended using HOOPRA-60. The author found it convenient to use HOOPRA-60 in a hands-free manner by wearing it over the shoulder. Assessing the population within a 5 km radius of JOFG, approximately 0.15 million individuals (61 thousand households) have the ability to listen to JOFG broadcasts using HOOPRA-60.

In the event of a major earthquake striking at night and causing a power outage, items in a room may become scattered and difficult to locate in complete darkness. Even if one possesses a disaster-prevention radio, finding it amidst the darkness can be challenging. However, HOOPRA-60, equipped with luminescent material having ultrahigh brightness, remains visible in such circumstances (Figure 1C). By hanging HOOPRA compactly on the walls near entrances of houses, it can be quickly identified and retrieved during a disaster. Additionally, HOOPRA-60 exhibits excellent visibility when worn over the shoulder, whether in evacuation centers or outdoors at night. The author was able to locate HOOPRA easily in the dark, and its hands-free usage while listening to JOFG broadcasts proved exceptionally valuable during disaster drills.

### 3.4. Field Experiments in Kanto Region

The author conducted an investigation to determine the extent to which JOAK and JOAB broadcasts could be comprehended using HOOPRA-60 and HOOPRA-80. As JOAK and JOAB transmit from the same location at Shobukuki, the straight-line distance from the site was considered. While the preliminary findings on reception limits emerged from research on radio-wave power generation, this paper focuses on the results of radio reception, with detailed research on radio-wave power generation to be reported separately. The author verified that HOOPRA-60 enables reception of both JOAK and JOAB within a radius of 50 km from Shobukuki. However, reception becomes challenging around the 100 km mark near Gotemba City, Shizuoka Prefecture. Based on these findings, the reception limit for HOOPRA-60 is estimated to fall between 60 km and 100 km. Since MW broadcasts primarily rely on ground waves for propagation within 100 km due to their low frequency, their transmission is influenced by factors such as the electrical conductivity of the ground, geological properties, and the terrain. NHK's broadcast area map indicates that the western boundary of the electric field strength of 3 mV/m extends to Nagano City and Kofu City, both of which are over 100 km away from Shobukuki [20]. In particular, no instances of anomalous propagation existed within a 100 km radius of Shobukuki, as the radio waves propagate nearly uniformly [24].

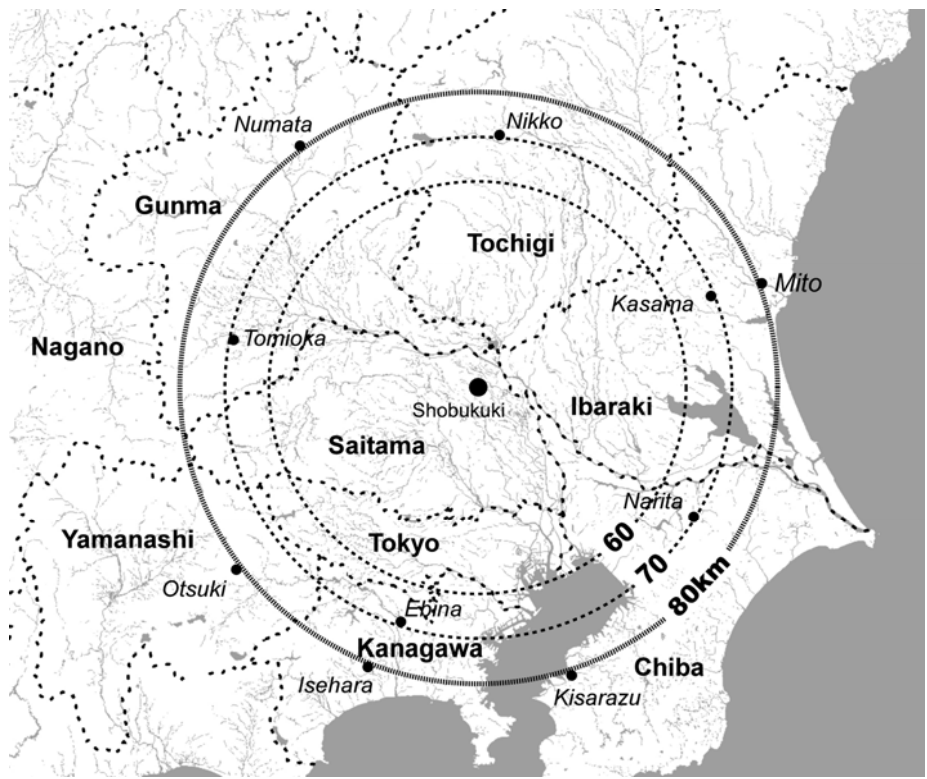

Figure 2   Map of Kanto region depicting circles of 60, 70, and 80 km radii from Shobukuki, corresponding to the location of NHK Tokyo Radio First Broadcast (JOAK) and Second Broadcast (JOAB) stations. Nonpowered HOOPRA-60 and HOOPRA-80 can be practically used within a 70 and 80 km radius, respectively. In a tranquil area, HOOPRA-60 barely managed to receive broadcasts in a 80 km radius.

At the 60 km point of Kita Yasaka Park, the induced voltage of HOOPRA-80 for JOAB was 70 mV, and the author could clearly understand the broadcast content of JOAK and JOAB using HOOPRA-60. Similarly, at the 70 km point of Kodomo Shizen Park, the induced voltage of HOOPRA-80 for JOAB was 30 mV. At the Hikichidai Park located at the 70 km mark, the induced voltage of HOOPRA-80 for JOAB was 30 mV. At both locations, the author could understand the broadcast content clearly using HOOPRA-60. The induced voltage remained the same at the 70 km mark in Kodomo Shizen Park.

Furthermore, at the 80 km point of Kanazawa Zoo, the induced voltage of HOOPRA-80 for JOAB was 20 mV. The author could comprehend the content of JOAK and JOAB broadcasts using HOOPRA-80. However, with HOOPRA-60, the author could only understand the broadcast content barely. Based on these findings, the author determined that the practical reception limit for JOAK and JOAB using HOOPRA-60 and HOOPRA-80 is approximately 70 km and 80 km, respectively. Upon analyzing the audio waveform of the recorded broadcast content and its audibility, it is determined that the broadcast closely resembled the results observed at the 15 km point in Fukui. The field experiments demonstrated that the author could receive JOAK and JOAB broadcasts within an 80 km radius from Shobukuki using HOOPRA-60 and HOOPRA-80 (Figure 2). Notably, the distance may vary slightly depending on the geographical conditions and geological characteristics. However, the reception distance can be extended by enhancing the sensitivity of the detection circuit or exploring antenna material improvements. Despite the difference in transmission power between JOAK (300 kW) and JOAB (500 kW), the author did not perceive a significant difference in reception quality. The slightly lower transmission frequency of JOAK, which is approximately 100 kHz lower than that of JOAB, may influence the propagation distance. The interaction between the voltage–current characteristics of the germanium diode in the circuit and small current of the earphone during reception could impact audibility.

## 4. Conclusion

Based on the field trials conducted in Fukui, the present study determined that HOOPRA-60 and HOOPRA-80 could be used for practical listening to JOFG broadcasts within a radius of approximately 13 km and 15 km from JOFG, respectively. In exceptionally quiet areas, even HOOPRA-60 managed to receive broadcasts within a 15 km radius, even though with limited clarity. In Fukui, an estimated 0.4 million people residing within a 15 km radius can utilize HOOPRA-80 to listen to the JOFG broadcast. JOFG represents one of the several 5 kW class radio broadcasting stations nationwide. Consequently, in most cities and towns where stations transmit at 5 kW power, HOOPRA-60 and HOOPRA-80 can be used within a radius of around 13 and 15 km from each station, respectively. For stations with higher transmission power than 5 kW, the broadcasting distance of 15 km radius will be increased. The field trials in the Kanto region revealed that HOOPRA-60 and HOOPRA-80 could enable reception of JOAK and JOAB broadcasts within a radius of approximately 70 km and 80 km from Shobukuki, respectively (Figure 2). Even in very quiet areas, HOOPRA-60 demonstrated reception capability within an 80 km radius, encompassing Saitama and Tokyo. This is significant because ~37.5 million people within an 80 km radius can listen to JOAK and JOAB broadcasts solely using the energy of radio waves with HOOPRA-80. As MW propagates primarily as ground waves, the destination distance is subject to variations based on geographical and geological factors. Note that HOOPRA cannot be utilized nationwide owing to radio-engineering principles and physical phenomena considerations. However, in regions with MW radio broadcasting stations or nationwide relay stations, HOOPRA can function as a disaster-prevention radio. Furthermore, even when the induced voltage was dropped below 20 mV, connecting HOOPRA-60 to a smartphone via a dedicated cable facilitated enhanced clarity of the broadcasted content.

The lessons learned from the Great Kanto Earthquake remind us of the importance of swiftly obtaining accurate information during a disaster and limiting the propagation of rumors. Even in modern times, a century after the Great Kanto Earthquake, we must remember the lessons related to the usage of smartphones and internet in such scenarios. As such, the radio can serve as a rapid and reliable source of information in the event of a disaster because the internet and television may not be available then due to power outages. Even in case of a power outage at night during a disaster, we can realize a method that enables over 30 million individuals in the Kanto region to receive radio broadcasts of JOAK or JOAB. The significance of MW broadcasting should be re-recognized precisely because of Japan's frequent natural disasters. Although MW commercial broadcasting is at the brink of operational shutdown because of business challenges, the potential of MW public broadcasting for saving lives in times of disaster should be reaffirmed to incorporate it into nationwide disaster-prevention infrastructure. The author believes that this research will contribute to such discussion. Moreover, as several high-output transmitting stations are functional worldwide, nonpowered HOOPRA technology can contribute to disaster-prevention applications and educational programs. Considering this centenary anniversary as an opportunity along with the numerous practical advantages of MW broadcasting, the author proposes that this vital infrastructure should be redeveloped as a matter of national policy to protect public safety and security.

## Acknowledgments

The author sincerely thanks his wife, Naoko Shoji, for her invaluable assistance in the transportation of equipment, experimental support, and valuable discussions throughout the field trials.

## References

[1] Cabinet Office, "Disaster Prevention Related Information – Reports Part 1 (1923 Great Kanto Earthquake)," Subcommittee of "Special Board of Inquiry on Inheriting the Lessons of Past Disasters" 2006.7
https://www.bousai.go.jp/kyoiku/kyokun/kyoukunnokeishou/rep/1923_kanto_daishinsai/index.html [Accessed May 9, 2023]

[2] Cabinet Office, "Disaster Prevention Related Information – Reports Part 2 (1923 Great Kanto Earthquake)," Subcommittee of "Special Board of Inquiry on Inheriting the Lessons of Past Disasters" 2008.3
https://www.bousai.go.jp/kyoiku/kyokun/kyoukunnokeishou/rep/1923_kanto_daishinsai_2/index.html [Accessed May 9, 2023]

[3] Cabinet Office, "Disaster Prevention Related Information – Reports Part 3 (1923 Great Kanto Earthquake)," Subcommittee of "Special Board of Inquiry on Inheriting the Lessons of Past Disasters" 2008.3
https://www.bousai.go.jp/kyoiku/kyokun/kyoukunnokeishou/rep/1923_kanto_daishinsai_3/index.html [Accessed May 9, 2023]

[4] Motoroni Kato, "Establishment of the Broadcasting System and Tsuyoshi Inukai: Analysis of Ministry of Communications internal materials and Imperial Diet responses", The NHK monthly report on broadcast research, 2011.4, pp58-69

[5] Hoso gojyunenshi [Honhen], edited by Japan Broadcasting Corporation, Japan Broadcasting Publishing Association 1977

[6] Hoso gojyunenshi [shiryohen], edited by Japan Broadcasting Corporation, Japan Broadcasting Publishing Association 1977

[7] Nihon Hososhi [Jokan](aka 35 nenshi), edited by Japan Broadcasting Corporation, Japan Broadcasting Publishing Association 1965

[8] Nihon Hososhi [Bekkan](aka 35 nenshi), edited by Japan Broadcasting Corporation, Japan Broadcasting Publishing Association 1965

[9] Nihon Hososhi (aka 25 nenshi), edited by Japan Broadcasting Corporation, Japan Broadcasting Corporation 1951

[10] The History of Broadcasting in Japan in the 20th Century [Jo-kan], edited by Japan Broadcasting Corporation, Japan Broadcasting Corporation 2001

[11] Guglielmo Marconi, "Wireless Telegraphic Communication: Nobel Lecture, 11 December 1909." Nobel Lectures. Physics 1901-1921. Amsterdam: Elsevier Publishing Company, 1967: pp196-222.

[12] John Grant, "Experiments and Results in Wireless Telephony", The American Telephone Journal, January 26, 1907, pp49-51:

[13] Anne F. MacLennan, "Celebrating a Hundred Years of Broadcasting – An Introduction and Timeline", Journal of Radio & Audio Media, 2020, Vol. 27-2, pp191–207

[14] Mamiko Naka, "The Voices and Hearts of the People were Broadcast from Radio, the Instrument of Civilization : Focusing on the Analysis of Periodicals in Outer Dalian and Inland Osaka", Journal of comprehensive cultural studies, 2022, Vol.16, pp22-60

[15] Miwako Nakamura, A Study of the 'National School Hour' Radio Programs, 1941-1945, Focusing on Scripts for Pre-schoolers, The Japanese Journal of the historical studies of early childhood education and care, 2019, 14, pp1-14

[16] Tsuguo Endo, Newly Boosted NHK Tokyo (Shobu Kuki) Radio Broadcasting Station, The Journal of the Institute of Television Engineers of Japan, 1982, Vol.36-9 pp841-843

[17] Statistics Bureau, Ministry of Internal Affairs and Communications, Population Estimates (as of October 1, 2022),
https://www.stat.go.jp/data/jinsui/2022np/index.html [Accessed May 9, 2023]

[18] Eiichi Shoji, "Magnetic Field Flexible Energy Harvester", Japanese Patent Application No. 2018-241880, Japanese Patent Application No. 2020-563278, PCT/JP2019/050477

[19] Eiichi Shoji, Development and Performance of a Battery-Free Disaster Prevention Radio "HOOPRA" Using the Energy Harvested from Radio Waves, *J. Disaster Res.* 2016, 11-3, 593-598

[20] Eiichi Shoji, "Multi-loop Type Antenna and Loop Antenna Type Passive Radio,, Japanese Patent Application No. 2021-163133

[21] Guidelines for consideration of disaster prevention measures in preparation for various forms of Nankai Trough earthquakes [1st edition], Cabinet Office (Disaster Management Section), Nankai Trough earthquake disaster prevention measures, Central Disaster Prevention Council,
https://www.bousai.go.jp/jishin/nankai/index.html [Accessed May 9, 2023]

[22] Information and Communications White Paper 2012, "Chapter 3 Lessons learned from the Great East Japan Earthquake and the role of ICT", pp255-287
https://www.soumu.go.jp/johotsusintokei/whitepaper/ja/h24/pdf/24honpen.pdf　[Accessed May 9, 2023]

[23] Statistics Bureau, Ministry of Internal Affairs and Communications, jSTAT MAP, Data: 2020 national census, https://www.e-stat.go.jp/

[24] Ministry of Internal Affairs and Communications AM and FM broadcasting
https://www.soumu.go.jp//000216453.pdf [Accessed May 9, 2023]